\RequirePackage{fix-cm}
\documentclass[smallextended]{svjour3}       
\smartqed  
\usepackage{lineno,hyperref}
\usepackage{graphicx,psfrag,epsf}
\usepackage{enumerate}
\usepackage{url} 
\usepackage{amsmath}
\usepackage{amsfonts}
\usepackage{multicol}
\usepackage{xfrac}
\usepackage{natbib}
\begin{document}

\title{Advanced Algorithms for Penalized Quantile and Composite Quantile Regression\thanks{Drs. Linglong Kong, Bei Jiang, and Di Niu are supported in part by the Natural Sciences and Engineering Research Council of Canada (NSERC).}
}


\author{Matthew Pietrosanu         \and
        Jueyu Gao \and
        Linglong Kong \and
        Bei Jiang \and
        Di Niu
}


\institute{M. Pietrosanu, J. Gao, L. Kong, and B. Jiang \at
              Department of Mathematical \& Statistical Sciences, University of Alberta, Edmonton, AB, Canada \\
              Tel.: +780-492-3396\\
              Fax: +780-492-6826\\
              \email{pietrosa@ualberta.ca, jueyu@ualberta.ca, lkong@ualberta.ca, bei1@ualberta.ca}
              \and
          D. Niu \at
              Department of Electrical and Computer Engineering, University of Alberta, Edmonton, AB, Canada \\
              Tel.: + 780-492-3332 \\
              \email{dniu@ualberta.ca}
             }

\date{Accepted: June 22, 2020, to appear in \emph{Computational Statistics}}

\maketitle

\begin{abstract}
In this paper, we discuss a family of robust, high-dimensional regression models for quantile and composite quantile regression, both with and without an adaptive lasso penalty for variable selection. We reformulate these quantile regression problems and obtain estimators by applying the alternating direction method of multipliers (ADMM), majorize-minimization (MM), and coordinate descent (CD) algorithms. Our new approaches address the lack of publicly available methods for (composite) quantile regression, especially for high-dimensional data, both with and without regularization. Through simulation studies, we demonstrate the need for different algorithms applicable to a variety of data settings, which we implement in the \texttt{cqrReg} package for R. For comparison, we also introduce the widely used interior point (IP) formulation and test our methods against the IP algorithms in the existing \texttt{quantreg} package. Our simulation studies show that each of our methods, particularly MM and CD, excel in different settings such as with large or high-dimensional data sets, respectively, and outperform the methods currently implemented in \texttt{quantreg}. The ADMM approach offers specific promise for future developments in its amenability to parallelization and scalability.
\end{abstract}

\section{Introduction}

With recent rising interest in sparse regression for high-dimensional data, least squares regression with regularization---often via lasso penalty \citep{lasso}---has become a focal point of computing scientists and statisticians in model selection procedures \citep{KongGenetic,L1}. Furthermore, quantile regression has emerged as an alternative to traditional ordinary least squares methods with numerous advantages, including but not limited to higher efficiency with heavy-tailed error distributions, robustness against outlying data, and more informative insights into the distribution of the response under study \citep{IP}.

Oracle model selection theory, introduced by \cite{FanLi}, illustrates optimal behaviour during model selection but is limited to the case where error variance is finite. In response, \cite{CQR} established composite quantile regression---a method to simultaneously model multiple quantile levels---that maintains desirable oracle properties even in the case of non-finite error variance. Beyond oracle model selection and the simultaneous modelling of multiple quantile levels, composite quantile regression also achieves a lower variance on estimated effects relative to quantile regression. These properties of composite quantile regression have proven attractive to many researchers who have widely applied this technique to improve the processing capabilities of artificial neural networks \citep{neural}, provide an alternative to local polynomial regression \citep{polyreg}, and smooth Harris chain stochastic processes \citep{harris}.

Applying existing optimization algorithms to (composite) quantile regression requires a non-trivial reformulation of the problem due to the non-linearity and non-differentiability of the loss and regularization terms of the objective function. The well-known \texttt{quantreg} package for R \citep{quantreg} uses an interior point (IP) approach for quantile and composite quantile regression, with native support for $l_1$ (lasso) regularization in only the former. Advanced IP algorithms in \texttt{quantreg}, e.g., using prediction-correction \citep{mehrotra} for non-regularized quantile regression, have greatly improved upon earlier simplex methods. However, the time spent on matrix inversion in IP approaches \citep{qrcomputation} motivates us to seek faster algorithms for quantile and composite quantile regression, particularly for high-dimensional data where regularization is required. \cite{adaptiveLasso}, following the conjectures of \cite{FanLi}, showed lasso variable selection---currently the most commonly implemented penalty for quantile regression---to be inconsistent in certain situations and presented adaptive lasso regularization as a solution. Our work in the present paper is thus motivated by both a search for faster quantile regression algorithms as well as the lack of publicly available methods for adaptive lasso regularized quantile and composite quantile regression, particularly for high-dimensional data.

Our work in this paper is novel in its approach to quantile regression, composite quantile regression, and corresponding versions regularized by an adaptive lasso penalty using three different algorithms. First, we present an alternating direction method of multipliers (ADMM) approach that breaks up the model estimation problem into simpler convex optimization problems that can be solved in parallel \citep{ADMMnew1}. Second, we give a majorize-minimization (MM) approach that iteratively minimizes a majorization, a particular differentiable approximation of the objective function containing both the quantile loss and penalty terms \citep{MM}. Third, we detail a coordinate descent (CD) method that uses observations in a greedy algorithm to iteratively select and update individual model parameters while holding others constant \citep{CD}. For the sake of comparison, we also discuss an IP formulation of the problem that seeks to minimize both loss and regularization functions after starting within rather than on the boundary of the feasible set \citep{IP}. In numerical simulations, we compare our approaches to the advanced IP methods present in the \texttt{quantreg} package. We implement the proposed methods using the publicly available \texttt{cqrReg} package for R \citep{Rpack}, which performs computations in C++ and links back to R via the \texttt{Rcpp} \citep{rcpp} and \texttt{RcppArmadillo} \citep{rcpparmadillo} packages for increased computational efficiency. The results of these simulations suggest that our approaches generally improve upon \texttt{quantreg}'s computation time with roughly the same level of estimation error for the range of quantile regression problems considered.  We find that the MM approach to non-regularized composite quantile regression greatly outperforms the other three methods in terms of computation time and that the CD method excels in regularized (composite) quantile regression with high-dimensional data. Our ADMM approach was at least comparable (in terms of computation time and estimate error) in most simulations performed but holds the promise of further improvement and scalability with distributed computing and parallelization. Indeed, ADMM has recently been explored in the context of penalized quantile regression for big data as well as in sparse settings \citep{yulin,admmhd}. Our new implementations provide users with new algorithms for quantile and composite quantile regression with competitive runtime in different data settings, all with comparable estimation error.

The rest of this article is structured as follows. Section \ref{sec:qr} presents quantile regression, starting with relevant notation in Subsection \ref{subsec:qrNotation}, followed by the description of our approaches to quantile regression using the ADMM, MM, and CD algorithms in Subsections \ref{subsec:qrADMM} through \ref{subsec:qrCD}. Section \ref{sec:cqr} continues with composite quantile regression, including relevant notation and commentary on the extension from quantile to composite quantile regression for our ADMM, MM, and CD methods. Numerical simulation results are presented in Section \ref{sec:sim} and discussed in Section \ref{sec:discussion}.

\section{Quantile Regression} \label{sec:qr}

In this section, we present the proposed ADMM, MM, and CD methods for quantile regression with adaptive lasso regularization. We refer interested readers to the online supplementary appendix for implementations of the non-regularized problems and further details (omitted for brevity) on our proposed methods. For completeness in the upcoming simulations, a basic IP formulation is also given in the online appendix.

\subsection{Background and Notation} \label{subsec:qrNotation}

We first introduce the necessary background and notation to be used throughout this paper regarding quantile regression, both with and without adaptive lasso regularization \citep{adaptiveLasso2,adaptiveLasso}. We are concerned with the linear model $$y = b_0 + \pmb x^T\pmb\beta + \varepsilon,$$ where we wish to estimate the level $\tau$ (for some $\tau\in(0,1)$) conditional quantile of $y\in\mathbb{R}$ given $\pmb x\in\mathbb{R}^p$, given by $b_0 + \pmb x^T\pmb\beta + b^\varepsilon_\tau$, where $b_\tau^\varepsilon$ is the (assumed unique) level $\tau$ quantile of the error distribution of $\varepsilon$, independent of $\pmb x$ \citep{CQR}.

For a fixed quantile level $\tau\in(0,1)$, define the quantile loss function, for any $t\in\mathbb{R}$, by $\rho_\tau(t) = \tau t_+ + (1-\tau)t_-$, where $t_+ = \max\{t,0\}$ and $t_- = \max\{-t,0\}$. Given a design matrix $\pmb X=[\pmb x_1|\dots|\pmb x_n]^T\in\mathbb{R}^{n\times p}$ and response variable vector $\pmb Y=(y_1,\dots,y_n)^T\in\mathbb{R}^n$, adaptive lasso regularized quantile regression estimates are obtained as $$(\hat b_0, \hat{\pmb \beta}) = \underset{b_0\in\mathbb{R},~\pmb\beta\in\mathbb{R}^p}{\arg\min}~\sum_{i=1}^n\rho_\tau(y_i - b_0 - \pmb x_i^T\pmb \beta) + p_{\lambda}(\lvert\pmb \beta\rvert),$$ where $\lambda>0$ is a regularization parameter, $p_{\lambda}(\lvert\pmb \beta\rvert)=\lambda\sum_{j=1}^p\sfrac{\lvert\beta_j\rvert}{\lvert{\beta_j^\text{QR}}\rvert^2}$ is the adaptive lasso penalty, and ${\pmb \beta}^\text{QR}=(\beta^\text{QR}_1,\dots,\beta^\text{QR}_p)^T\in\mathbb{R}^p$ is the estimator (without intercept) obtained from non-regularized quantile regression \citep{originalQR,IP}---that is, the estimator in the problem with $\lambda=0$.

Define the residuals for quantile regression by $r_i=r_i(b_0,\pmb \beta)=y_i-b_0-\pmb x_i^T\pmb \beta$, for $i=1,\dots,n$. For the ease of notation throughout this section, we sometimes assume that a design matrix $\pmb X$ has an appropriate column for the intercept term of the model. Where intercepts are accounted for in the design matrix, the parameter vector $\pmb \beta$ will be taken to include the corresponding intercept terms such that $\pmb \beta=(b_0,\beta_1,\dots,\beta_p)^T\in\mathbb{R}^{p+1}$. This will be made clear by the dimension of $\pmb \beta$. Throughout this paper, $p$ will always refer to the number of covariate parameters and $\beta_j$, for $j=1,\dots,p$, will always refer to a covariate effect and never an intercept term.

\subsection{Alternating Direction Method of Multipliers Algorithm} \label{subsec:qrADMM}

Although developed in the 1960s and 1970s \citep{ADMMhistory1,ADMMhistory2}, interest in the ADMM algorithm was renewed with the findings of \cite{ADMMnew1} and \cite{ADMMnew2}. These studies demonstrate the ADMM algorithm's relative efficiency in solving optimization problems with large data sets, particularly when non-smooth terms are present in the objective function. This method has found notable use in quantile regression where the quantile loss and regularization term (if present) are not differentiable \citep{ADMMnew1,ADMMkong1,ADMMkong2}. For brevity, a general formulation of the ADMM algorithm is available in the online supplementary appendix. We apply the ADMM algorithm \citep{ADMMnew1} by reformulating regularized quantile regression as the convex optimization problem
\begin{align*}
\underset{\pmb \beta\in\mathbb{R}^{p+1}}{\min}\qquad&\sum_{i=1}^n\rho_\tau(r_i)+ p_\lambda(\lvert\pmb \beta\rvert)\\
\text{subject to }\qquad&  \pmb X\pmb \beta+\pmb r=\pmb Y,
\end{align*}
where $\pmb r$ is a vector of residuals and where the intercept term is accounted for in both $\pmb \beta$ and $\pmb X$. We solve this problem using the ADMM iteration scheme \citep{ADMMnew1}
\begin{align*}
\pmb r^{(t+1)} &= \underset{\pmb r\in\mathbb{R}^n}{\arg\min}~~\sum_{i=1}^n\rho_\tau(r_i)+\frac{\rho}{2}||\pmb Y-\pmb r-\pmb X\pmb\beta^{(t)}+\pmb u^{(t)}/\rho||_2^2\\
\pmb \beta^{(t+1)} &= \underset{\pmb \beta\in\mathbb{R}^{p+1}}{\arg\min}~~\frac{\rho}{2}||\pmb Y-\pmb r^{(t+1)}-\pmb X\pmb \beta+\pmb u^{(t)}/\rho||_2^2+p_\lambda(\lvert\pmb \beta\rvert)\\
\pmb u^{(t+1)} &= \pmb u^{(t)} + \rho(\pmb Y-\pmb r^{(t+1)}-\pmb X\pmb \beta^{(t+1)}),
\end{align*}
where $\pmb u$ is the rescaled Lagrange multiplier and $\rho>0$ is a penalty parameter. For reference, $\rho$ is chosen to be 1.2 by \cite{ADMMnew1}. The update for $\pmb r$ can be written in a closed form as $S_{1/\rho}\big(\pmb c-(2\pmb \tau_{n\times1}-\pmb 1_{n\times1})/\rho\big)$ where $\pmb c=\pmb Y-\pmb X\pmb \beta^{(t)}+\pmb u^{(t)}/\rho$ and, for $a\in\mathbb{R}$, the soft thresholding operator $S_a:\mathbb{R}^m\rightarrow\mathbb{R}^m$ is defined component-wise via $(S_a(\pmb v))_i = (v_i-a)_+-(-v_i-a)_+$. Similarly, the update for $\pmb \beta$ does not have a closed form but can be viewed as a least squares optimization problem with adaptive lasso penalty. We implement existing numerical methods to solve this problem and update $\pmb \beta$.

Let $\pmb X_*$ and $\pmb \beta_*$ be $\pmb X$ and $\pmb \beta$ with the intercept term removed and $\pmb b$ a vector of intercepts $(b_0)_{n\times1}$. A generic stopping condition for the algorithm can be defined in terms of the primal and dual residuals $\pmb r_\text{primal}^{(t+1)}$ and $\pmb r_\text{dual}^{(t+1)}$, respectively, with the stopping conditions $\lvert\lvert \pmb r_\text{primal}^{(t+1)}\rvert\rvert_2\leq\varepsilon_\text{primal}$ and $\lvert\lvert \pmb r_\text{dual}^{(t+1)}\rvert\rvert_2\leq\varepsilon_\text{dual}$. In this regularized setting, we have (from the general ADMM algorithm) that
\begin{align*}
\pmb r_\text{primal}^{(t+1)} &= \pmb Y -\pmb X\pmb \beta^{(t+1)}- \pmb r^{(t+1)}\\
\pmb r_\text{dual}^{(t+1)} &= \rho \pmb X_*^T(\pmb r^{(t+1)}-\pmb r^{(t)})\\
\varepsilon_\text{primal} &= \sqrt{n}\varepsilon_\text{abs} + \varepsilon_\text{rel}\max\{\lvert\lvert \pmb X_*\pmb \beta_*^{(t+1)}\rvert\rvert_2^2, \lvert\lvert \pmb r^{(t+1)}\rvert\rvert_2^2, \lvert\lvert \pmb b-\pmb Y \rvert\rvert_2^2\},\\
\varepsilon_\text{dual} &= \sqrt{p}\varepsilon_\text{abs}+\varepsilon_\text{rel}\lvert\lvert \pmb X^T\pmb u^{(t+1)}\rvert\rvert_2^2,
\end{align*}
with possible tolerance values $\varepsilon_\text{abs}=10^{-4}$ and $\varepsilon_\text{rel}=10^{-2}$, respectively \citep{ADMMnew1}.

\subsection{Majorize-Minimization Algorithm} \label{subsec:qrMM}

The use of majorizing functions to solve minimization problems has been well-studied in the statistical literature for many years since \cite{ortegaCD}. It was not until a later time, however, that the general MM framework was put forward by \cite{MM}. In general, MM can refer to majorize-minimization or minorize-maximization, depending on whether the problem at hand is a minimization or maximization problem, respectively. MM algorithms operate iteratively by constructing an auxiliary function $g_t(\cdot|\pmb \beta^{(t)})$ using a solution $\pmb \beta^{(t)}$ for the current iteration that will simultaneously optimize the original objective function $f$. In the case of a minimization problem, this function is called a majorizer and must satisfy $g_t(\pmb \beta|\pmb \beta^{(t)})\geq f(\pmb \beta)$ for all $\pmb \beta$ of interest and $g_t(\pmb \beta^{(t)}|\pmb \beta^{(t)})=f(\pmb \beta^{(t)})$.
Arguably, the most well-known application of an MM method is in the expectation-maximization (EM) algorithm \citep{EM} for maximum likelihood estimation. MM has also been applied in various areas of research, e.g., regression, survival analysis, discriminant analysis, and quantile regression \citep{MMtutorial}. We use the MM algorithm developed by \cite{MM} and \cite{MMR} to solve the quantile regression problem with adaptive lasso regularization.

We first construct a function $\rho_\tau^\varepsilon(r)$ based on some perturbation parameter $\varepsilon>0$ to approximate the fidelity portion $\sum_{i=1}^n \rho_\tau(r_i)$ of the objective function. For any $r\in\mathbb{R}$, define $\rho_\tau^\varepsilon(r) = \rho_\tau(r) - \frac{\varepsilon}{2}\ln(\varepsilon+|r|)$ so that the fidelity can be approximated by $\sum_{i=1}^n\rho_\tau^\varepsilon(r_i)$. At the $t$-th iteration, for each residual value $r_i^{(t)}=r_i^{(t)}(\pmb \beta^{(t)})$, we have that $\rho_\tau^\varepsilon(r)$ is majorized by the quadratic function $$\xi_\tau^\varepsilon(r|r_i^{(t)}) = \frac{1}{4}\Biggl[\frac{r^2}{\varepsilon+|r_i^{(t)}|}+(4\tau-2)r+c\Biggr],$$ for some solvable constant $c$ that satisfies the equation $\xi(r_i^{(t)}|r^{(t)})=\rho_\tau^\varepsilon(r^{(t)})$. Given $\lambda$, $\pmb \beta^\text{QR}$, and an initial value $\pmb \beta^{(0)}=(\beta_1^{(0)},\dots,\beta_p^{(0)})$ for $\pmb \beta$, we can locally approximate the penalty $p_\lambda(\lvert\pmb \beta\rvert)$ as a quadratic function. This yields a majorizer of the objective function \citep{MMR}, $$Q^\varepsilon(\pmb \beta|\pmb \beta^{(t)}) = \sum_{i=1}^n\xi_\tau^\varepsilon(r_i|r_i^{(t)})+\lambda\sum_{j=1}^p\frac{1}{\lvert\beta^\text{QR}_j\rvert^2}\Biggl[|\beta^{(t)}_j|+ \frac{\big(\beta_j^2-(\beta^{(t)}_j)^2\big)\mathrm{sgn}(\beta_j^{(t)})}{2|\beta^{(t)}_j+\varepsilon|}\Biggr].$$ For the $t$-th iteration of the algorithm, given an updated value $\pmb \beta^{(t)}$ for $\pmb \beta$, we minimize the quadratic function $Q^\varepsilon(\cdot|\pmb \beta^{(t)})$ using a Newton-Raphson iterative method. The argument minimum is used to update $\pmb \beta^{(t)}$ and can be used to decide when to terminate the algorithm. For our purposes, we use tolerance $10^{-3}$.

\subsection{Coordinate Descent Algorithm} \label{subsec:qrCD}

Coordinate descent (CD) algorithms are iterative procedures that generally fix some components of the argument vector in an optimization problem and solve the resulting subproblem in terms of the unfixed components. CD methods have a long-standing history \citep{ortegaCD} and their convergence properties are well-documented \citep{earlyCD1,earlyCD2}. The simplest CD algorithms allow for exactly one unfixed variable per iteration and search for a subproblem solution along a line, while others will search along a hyperplane by allowing multiple unfixed components. Most implementations use the latter in a block coordinate descent method. CD methods have been developed extensively, particularly for non-differentiable, non-convex objective functions, permitting the use of regularization functions such as lasso ($l_1$) and ridge ($l_2$) penalties \citep{earlyCD2,CDtibshirani}.

To implement quantile regression with adaptive lasso regularization, we use an extended version of the greedy CD method put forward by Edgeworth and, more recently, further developed by \cite{CD}. This requires us to reformulate the quantile objective function. In each iteration, for fixed $\pmb \beta\in\mathbb{R}^p$, replace $b_0$ by the level-$\tau$ sample quantile of the residuals $y_i-\pmb X_i^T\pmb \beta$ for $i=1,\dots,n$: this will necessarily drive the value of the objective function downwards. Define $\Theta_i = \rho_\tau(r_i)$ for $i=1,\dots,n$. For $m=1,\dots,p$, rewrite the loss function as $$L(b_0,\pmb \beta)=L_{m}(b_0,\pmb \beta) = \sum_{i=1}^n\lvert x_{im}\rvert\Biggl\lvert\frac{y_i-b_0-\sum_{j=1,\,j\neq m}^px_{ij}\beta_j}{x_{im}}-\beta_m\Biggr\rvert\cdot\Theta_i+p_\lambda(\lvert\pmb \beta\rvert)$$ and apply the CD algorithm. For each fixed $m$, define $z_i = \frac{1}{x_{im}}\big(y_i-b_0-\sum_{j=1,\,j\neq m}^px_{ij}\beta_j\big)$ if $r_i\geq0$ and $z_i=0$ if $r_i<0$. We sort $z_i$, for $i=1,\dots,n$, and update $\beta_m$ to the value of the $i^*$-th order statistic $z_{(i^*)}$ satisfying 
$$\sum_{j=1}^{i^*-1}w_{(j)} < \frac{1}{2}\sum_{j=1}^n w_{(j)}
\qquad\text{and}\qquad
\sum_{j=1}^{i^*}w_{(j)} \geq \frac{1}{2}\sum_{j=1}^n w_{(j)},$$
where $w_i=\lvert x_{im}\rvert\cdot\theta_i$ if $r_i\geq0$ and $w_i=\sfrac{\lambda}{\lvert\beta_m^\text{QR}\rvert^2}$ if $r_i<0$. In other words, using the weights $w_i$, the selected $z_{(i^*)}$ is the weighted median of all $z_i$ (for the fixed value of $m$). At the end of each iteration, check for the convergence of $\pmb \beta$ using the selected stopping criteria. Here, we use an absolute value difference threshold of $10^{-3}$.

\section{Composite Quantile Regression} \label{sec:cqr}

In this section, we present an extension from quantile to composite quantile regression for the proposed ADMM, MM, and CD algorithms. We only show results for the case with adaptive lasso regularization. Readers interested in the non-regularized case are referred to the online supplementary appendix where more details and a similar extension for a basic IP formulation are given. With regards to the available \texttt{quantreg} package for R \citep{quantreg}, we note that non-regularized composite quantile regression has only recently been implemented using an IP algorithm and that a regularized version is currently not natively available without further reformulation of the problem.

Composite quantile regression \citep{CQR} simultaneously estimates a sequence of $K$ conditional quantiles of $y$ given $\pmb X$ at levels $0<\tau_1<\tau_2<\dots<\tau_K<1$. Under the same linear model as before, these conditional quantiles are given by $b_0 + \pmb X^T\pmb\beta + b^\varepsilon_k$, where $b^\varepsilon_k$ is the (assumed unique) level $\tau_k$ quantile of the error distribution of $\varepsilon$, again assumed independent to be independent of $\pmb X$. Unlike $K$ independent quantile regression models, the composite model assumes the same covariate effects across the $K$ quantile levels. Adaptive lasso regularized composite quantile regression estimates are obtained as $$(\hat{b}_1,\dots,\hat{b}_K,\hat{\pmb \beta}^\text{CQR}) = \underset{b_1,\dots,b_K\in\mathbb{R},~\pmb\beta\in\mathbb{R}^p}{\arg\min}~\sum_{k=1}^K\sum_{i=1}^n \rho_\tau(y_i - b_k - \pmb X_i^T\pmb \beta)+p_\lambda(\lvert\pmb \beta\rvert),$$ where $\lambda>0$ is a regularization parameter, $p_\lambda(\lvert\pmb \beta\rvert)=\lambda\sum_{j=1}^p\sfrac{\lvert\beta_j\rvert}{\lvert\beta_j^\text{CQR}\rvert^2}$, and $\pmb \beta^\text{CQR}$ is the solution (without intercepts) to the non-regularized composite quantile regression problem. To extend the residual notation defined before, let $r_{ik}=y_i-b_k-\pmb x_i^T\pmb \beta$, for $i=1,\dots,n$ and $k=1,\dots,K$. \cite{CQR} impose regularity conditions to ensure the asymptotic normality of the unregularized composite quantile estimates: the authors note these are essentially the same as those in standard quantile regression \cite{IP}.

The extension from quantile to composite quantile regression is relatively straightforward: we need only accommodate additional quantile levels and intercept terms. Since the composite quantile case only adds more intercept parameters, the penalty term remains unchanged. For explicit details on our methods for regularized composite quantile regression in the ADMM, MM, and CD approaches, refer to the online supplementary appendix.

To extend the ADMM method, we generate a new design matrix $\pmb X^*\in\mathbb{R}^{nK\times (p+K)}$ by ``stacking'' the design matrices for each quantile level and adjusting all input accordingly. Written formally,
\begin{center}
$\pmb X^*_{nK\times(p+K)}=
\begin{bmatrix}
			[\pmb 1&\pmb 0&\pmb 0&\cdots&\pmb 0]&\pmb X\\
			[\pmb 0&\pmb 1&\pmb 0&\cdots&\pmb 0]&\pmb X\\
			[\pmb 0&\pmb 0&\pmb 1&\cdots&\pmb 0]&\pmb X\\
			\vdots&\vdots&\vdots&\vdots&\vdots&\vdots\\
			[\pmb 0&\pmb 0&\pmb 0&\cdots&\pmb 1]&\pmb X\\
\end{bmatrix}$, 
$\pmb Y^*_{nK\times1}=
\begin{bmatrix}
		\pmb Y\\
		\pmb Y\\
		\pmb Y\\
		\vdots\\
		\pmb Y
		\end{bmatrix}$,
$\pmb b^*=
\begin{bmatrix}
		{(b_1)}_{n\times1}\\
		{(b_2)}_{n\times1}\\
		{(b_3)}_{n\times1}\\
		\vdots\\
		{(b_K)}_{n\times1}\\
\end{bmatrix}$,
$\pmb \tau^*=
\begin{bmatrix}
		{(\tau_1)}_{n\times1}\\
		{(\tau_2)}_{n\times1}\\
		{(\tau_3)}_{n\times1}\\
		\vdots\\
		{(\tau_K)}_{n\times1}\\
\end{bmatrix}$,
\end{center}
where, for example, $[\pmb 1~\pmb 0~\pmb 0~\cdots~\pmb 0]$ denotes the $n\times K$ matrix with rows $(1,~0,~\cdots~,0)^T\in\mathbb{R}^K$. The methods presented in Subsection \ref{subsec:qrADMM} for quantile regression then apply after replacing $\pmb X$, $\pmb Y$, $\pmb b$, and $\tau$ with $\pmb X^*$, $\pmb Y^*$, $\pmb b^*$, and $\pmb \tau^*$, respectively. After replacement, the optimization problem becomes
\begin{align*}
\underset{\beta\in\mathbb{R}^{p+K}}{\min}\qquad&\sum_{k=1}^K\sum_{i=1}^n\rho_{\tau_k}(r_{ik})+ p_\lambda(\lvert\pmb \beta\rvert)\\
\text{subject to }\qquad&  \pmb X^*\pmb \beta+\pmb r=\pmb Y^*.
\end{align*}
With these changes, the explicit update scheme for ADMM is given by
\begin{align*}
\pmb r^{(t+1)} &= \underset{\pmb r\in\mathbb{R}^{nK}}{\arg\min}~~\sum_{k=1}^K\sum_{i=1}^n\rho_{\tau_k}(r_{ik})+\frac{\rho}{2}||\pmb Y^*-\pmb r-\pmb X^*\pmb \beta^{(t)}+\pmb u^{(t)}/\rho||_2^2\\
\pmb \beta^{(t+1)} &= \underset{\pmb \beta\in\mathbb{R}^{p+K}}{\arg\min}~~\frac{\rho}{2}||\pmb Y^*-\pmb r^{(t+1)}-\pmb X^*\pmb \beta+\pmb u^{(t)}/\rho||_2^2+\lambda\sum_{j=1}^p\sfrac{\lvert{\beta}_j\rvert}{\lvert\beta_j^\text{CQR}\rvert^2}\\
\pmb u^{(t+1)} &= \pmb u^{(t)} + \rho(\pmb Y^*-\pmb r^{(t+1)}-\pmb X^*\pmb \beta^{(t+1)}),
\end{align*}
where $\pmb c=\pmb Y^*-\pmb X^*\pmb \beta^{(t)}+\pmb u^{(t)}/\rho$, with residuals
\begin{align*}
\pmb r_\text{primal}^{(t+1)} &= \pmb Y^* -\pmb X^*\pmb \beta^{(t+1)}- \pmb r^{(t+1)}\\
\pmb r_\text{dual}^{(t+1)} &= \rho {\pmb X_*^*}^T(\pmb r^{(t+1)}-\pmb r^{(t)})\\
\varepsilon_\text{primal} &= \sqrt{n}\varepsilon_\text{abs} + \varepsilon_\text{rel}\max\{\lvert\lvert \pmb X_*^*\pmb \beta_*^{(t+1)}\rvert\rvert_2^2, \lvert\lvert \pmb r^{(t+1)}\rvert\rvert_2^2, \lvert\lvert \pmb b^*-\pmb Y \rvert\rvert_2^2\},\\
\varepsilon_\text{dual} &= \sqrt{p}\varepsilon_\text{abs}+\varepsilon_\text{rel}\lvert\lvert {\pmb X^*}^T\pmb u^{(t+1)}\rvert\rvert_2^2.
\end{align*}

The extension of the remaining two methods is similar, although requiring a slight change in the objective function. For the CD method, we modify our reformulation $L_m$ of the objective function, for $m=1,\dots,p$, to include a second summation for the additional quantile levels as $$L_m(b_1,\dots,b_k,\pmb \beta) = \sum_{k=1}^K\sum_{i=1}^n\lvert x_{im}\rvert\Biggl\lvert\frac{y_i-b_k-\sum_{j=1,\,j\neq m}^px_{ij}\beta_j}{x_{im}}-\beta_m\Biggr\rvert\cdot\Theta_{ik}+p_\lambda(\lvert\pmb \beta\rvert),$$ where $\Theta_{ik} = \rho_{\tau_k}(r_{ik})$ is analogous to $\Theta_i$ defined previously. The MM approach is similarly extended, yielding a final majorizer of the form $$Q^\varepsilon(\pmb \beta|\pmb \beta^{(t)}) = \sum_{k=1}^K\sum_{i=1}^n\xi_{\tau_k}^\varepsilon(r_{ik}|r_{ik}^{(t)})+\lambda\sum_{j=1}^p\frac{1}{\lvert\beta_j^\text{CQR}\rvert^2}\Biggl[|\beta^{(t)}_j|+ \frac{\big(\beta_j^2-(\beta^{(t)}_j)^2\big)\mathrm{sgn}(\beta_j^{(t)})}{2|\beta^{(t)}_j+\varepsilon|}\Biggr].$$

\section{Numerical Simulations} \label{sec:sim}

In this section, we evaluate the performance of the proposed ADMM, MM, and CD methods against that of the IP methods in \texttt{quantreg}. Because \texttt{quantreg} does not natively support regularized composite quantile regression, we do not make a comparison with IP approaches in that setting. Lasso regularization is used in place of adaptive lasso regularization for the IP method as the latter is not readily available in \texttt{quantreg}. Throughout this section, data is generated according to the model $$y_i = b+\sum_{j=1}^px_{ij}\beta_j+\varepsilon_i,$$ for $i=1,\dots,n$, where the $\varepsilon_i$ are i.i.d. standard normal random variables. We use a convergence threshold of $10^{-4}$ to define our stopping criteria throughout.

We first focus on parameter estimation rather than variable selection and consider cases with $p=5$ variables and $n=200,400,600,800,1000,2000$ observations in non-regularized quantile and composite quantile regression. In each simulation, the true value of each $\beta_j$ is uniform randomly sampled from the interval $[-1,1]$. In the quantile regression case, we set $\tau=0.3$ and in the composite quantile setting, we use quantile levels $0.1,0.2,\dots,0.9$. Tables \ref{tab:qr} and \ref{tab:cqr} present the performance of each method, averaged over 50 simulations.

\begin{table}[h]
\centering
\begin{tabular}{|l|ll|ll|ll|ll|}
\hline
\multicolumn{1}{|c|}{($n$,$p$)} & \multicolumn{2}{c|}{IP}                               & \multicolumn{2}{c|}{ADMM} & \multicolumn{2}{c|}{MM}                               & \multicolumn{2}{c|}{CD}                               \\ \hline
                            & \multicolumn{1}{c}{Error} & \multicolumn{1}{c|}{Time} & Error       & Time        & \multicolumn{1}{c}{Error} & \multicolumn{1}{c|}{Time} & \multicolumn{1}{c}{Error} & \multicolumn{1}{c|}{Time} \\ \hline
(200,5)                     & 0.08                      & 0.002                     & 0.063       & 0.002       & 0.060                     & \textbf{0.0002}                    & \textbf{0.036}                     & 0.002                     \\
(400,5)                     & 0.052                     & 0.0022                    & 0.055       & 0.0038      & 0.051                     & \textbf{0.0004}                    & \textbf{0.046}                     & 0.003                     \\
(600,5)                     & 0.043                     & 0.0029                    & 0.042       & 0.005       & \textbf{0.033 }                    & \textbf{0.0005}                    & 0.043                     & 0.0416                    \\
(800,5)                     & 0.037                     & 0.0048                    & 0.035       & 0.006       & \textbf{0.031}                     & \textbf{0.0005}                    & 0.034                     & 0.0046                    \\
(1000,5)                    & 0.0336                    & 0.0053                    & 0.031       & 0.008       & \textbf{0.026}                     & \textbf{0.0006}                    & 0.031                     & 0.0064                    \\
(2000,5)                    & 0.0213                    & 0.01                      & 0.022       & 0.013       & \textbf{0.018}                     & \textbf{0.001}                     & 0.022                     & 0.0096                    \\ \hline
\end{tabular}
\caption{Simulation results for quantile regression without regularization. Time measures the average computation time in seconds over 50 replications and Error measures the average absolute value difference between the estimated and true parameter values. The IP column displays the results from quantile regression using the IP method available in \texttt{quantreg}. The lowest Error and Time values for each ($n$,$p$) are noted in bold.}\label{tab:qr}
\end{table}

\begin{table}[h]
\centering
\begin{tabular}{|l|ll|ll|ll|ll|}
\hline
\multicolumn{1}{|c|}{($n$,$p$)} & \multicolumn{2}{c|}{IP}                               & \multicolumn{2}{c|}{ADMM} & \multicolumn{2}{c|}{MM}                               & \multicolumn{2}{c|}{CD}                               \\ \hline
                            & \multicolumn{1}{c}{Error} & \multicolumn{1}{c|}{Time} & Error       & Time        & \multicolumn{1}{c}{Error} & \multicolumn{1}{c|}{Time} & \multicolumn{1}{c}{Error} & \multicolumn{1}{c|}{Time} \\ \hline
(200,5)                     & 0.058                     & 0.009                     & \textbf{0.057}       & 0.029       & 0.057                     & \textbf{0.0008}                    & 0.058                     & 0.008                     \\
(400,5)                     & 0.043                     & 0.021                     & 0.043       & 0.057       & 0.047                     & \textbf{0.001}                     & \textbf{0.040}                      & 0.011                     \\
(600,5)                     & 0.035                     & 0.03                      & \textbf{0.034}       & 0.088       & 0.034                     & \textbf{0.0012}                    & 0.039                     & 0.017                     \\
(800,5)                     & 0.029                     & 0.047                     & \textbf{0.029}       & 0.122       & 0.029                     & \textbf{0.0014}                    & 0.031                     & 0.018                     \\
(1000,5)                    & 0.025                     & 0.064                     & \textbf{0.024}       & 0.16        & 0.028                     & \textbf{0.0015}                    & 0.024                     & 0.025                     \\
(2000,5)                    & 0.077                     & 0.14                      & \textbf{0.017}       & 0.36        & 0.017                     &\textbf{0.0026}                    & 0.018                     & 0.044                     \\ \hline
\end{tabular}
\caption{Simulation results for composite quantile regression without regularization. Time measures the average computation time in seconds over 50 replications and Error measures the average absolute value difference between the estimated and true parameter values. The IP column displays results from composite quantile regression using the IP method available in \texttt{quantreg}. The lowest Error and Time values for each ($n$,$p$) are noted in bold.}\label{tab:cqr}
\end{table}

We next consider variable selection for high-dimensional data using $n=100,200,500$ and varying $p$ from $1.5n$ to $5n$. The performance of each algorithm is summarized by the average number of false predictors selected, the average number of true predictors selected, and the average computation time in seconds over 25 replications. Simulation results in Table \ref{table:1} are for regularized quantile regression with quantile level $\tau=0.3$: here, the ADMM, MM, and CD methods use adaptive lasso regularization as described in previous sections, while the IP method uses the lasso regularization available in \texttt{quantreg}. Table \ref{table:2} gives results based on composite quantile regression with adaptive lasso regularization using quantile levels $0.1, 0.2, \dots, 0.9$: we do not make a comparison against an IP approach here, however, as a comparable method is not readily available in \texttt{quantreg}.

\begin{table}[h]
\setlength\tabcolsep{4pt} 
\centering
\begin{tabular}{|l|lll|lll|lll|lll|}
\hline
\multicolumn{1}{|c|}{($n$,$p$)} & \multicolumn{3}{c|}{IP} & \multicolumn{3}{c|}{ADMM} & \multicolumn{3}{c|}{MM} & \multicolumn{3}{c|}{CD} \\ \hline
                                & Time   & $N_T$  & $N_F$ & Time    & $N_T$  & $N_F$  & Time   & $N_T$  & $N_F$ & Time   & $N_T$  & $N_F$ \\ \hline
(100,200)                       & 0.074  & 4      & 0     & 0.017   & 4      & 0      & 0.1    & 4      & 0.1   & \textbf{0.014}  & 4      & 0     \\
(100,300)                       & 0.024  & 4      & 0     & 0.041   & 4      & 0      & 0.25   & 4      & 0     & \textbf{0.02}   & 4     & 0     \\
(100,500)                       & 0.98   & 4      & 0     & 0.152   & 4      & 0      & 0.812  & 3.9    & 0     & \textbf{0.035}  & 4      & 0     \\
(200,400)                       & 0.627  & 4      & 0     & 0.088   & 4      & 0      & 0.58   & 4      & 0     & \textbf{0.048}  & 4      & 0     \\
(200,600)                       & 1.96   & 4      & 0     & 0.161   & 4      & 0      & 1.64   & 4      & 0     & \textbf{0.054}  & 4      & 0     \\
(200,1000)                      & 8.85   & 4      & 0     & 0.791   & 4      & 0      & 6.23   & 4      & 0     & \textbf{0.11}   & 4      & 0     \\
(500,750)                       & 5.1    & 4      & 0     & 0.522   & 4      & 0      & 4.09   & 4      & 0     & \textbf{0.18}   & 4      & 0     \\
(500,1000)                      & 11     & 4      & 0     & 0.852   & 4      & 0      & 10.3   & 4      & 0     & \textbf{0.24}   & 4      & 0     \\
(500,1500)                      & 38     & 4      & 0     & 2.41    & 4      & 0      & 24     & 4      & 0     & \textbf{0.36}   & 4      & 0     \\ \hline
\end{tabular}
\caption{Simulation results for regularized quantile regression: the ADMM, MM, and CD methods use adaptive lasso, while the IP method from \texttt{quantreg} uses lasso regularization. Time measures the average computation time in seconds over 25 replications; $N_T$ and $N_F$ give the average number of true and false predictors selected, respectively. The lowest Time value for each ($n$,$p$) is noted in bold.}\label{table:1}
\end{table}


\begin{table}[h]
\centering
\begin{tabular}{|l|lll|lll|lll|}
\hline
\multicolumn{1}{|c|}{($n$,$p$)} & \multicolumn{3}{c|}{ADMM} & \multicolumn{3}{c|}{MM} & \multicolumn{3}{c|}{CD} \\ \hline
                                & Time    & $N_T$  & $N_F$  & Time  & $N_T$  & $N_F$  & Time   & $N_T$  & $N_F$ \\ \hline
(100,200)                       & \textbf{0.043}   & 4      & 0      & 0.11  & 4      & 0.8    & 0.13   & 4      & 0     \\
(100,300)                       & \textbf{0.089}   & 4      & 0      & 0.29  & 4      & 0.6    & 0.18   & 4      & 0     \\
(100,500)                       & \textbf{0.21}    & 4      & 0      & 1.01  & 4      & 0.64   & 0.32   & 4      & 0     \\
(200,400)                       & \textbf{0.22}    & 4      & 0      & 0.75  & 4      & 0.64   & 0.47   & 4      & 0     \\
(200,600)                       & \textbf{0.452}   & 4      & 0      & 1.9   & 4      & 0.72   & 0.676  & 4      & 0     \\
(200,1000)                      & 1.41    & 4      & 0      & 7.4   & 4      & 0.25   & \textbf{0.615}  & 4      & 0     \\
(500,750)                       & \textbf{1.52}    & 4      & 0      & 5.4   & 4      & 0.8    & 2.4    & 4      & 0     \\
(500,1000)                      & \textbf{2.43}    & 4      & 0      & 10.3  & 4      & 0.8    & 2.6    & 4      & 0     \\
(500,1500)                      & 5.86    & 4      & 0      & 28.5  & 4      & 0      & \textbf{3.7}    & 4      & 0     \\ \hline
\end{tabular}
\caption{Simulation results for composite quantile regression with adaptive lasso regularization for the ADMM, MM, and CD algorithms. An IP method from \texttt{quantreg} is not available in this setting. Time measures the average computation time in seconds over 25 replications; $N_T$ and $N_F$ give the average number of true and false predictors selected, respectively.}\label{table:2}
\end{table}

\section{Discussion and Conclusions}\label{sec:discussion}

In this paper we have presented three novel approaches to quantile and composite quantile regression and variable selection. Motivated by the lack of variety in algorithms for (composite) quantile regression, both with and without adaptive lasso regularization, and a desire to improve run times over the existing IP methods, we reformulated four types of quantile regression problems and presented estimators obtained using three algorithms. Using our existing implementation of these methods in the \texttt{cqrReg} package for R \citep{Rpack}, we used simulation studies to compare our methods to the IP algorithms available in the \texttt{quantreg} package \citep{quantreg}.

In the non-regularized quantile regression setting, we do not observe substantial differences in the average estimation error between methods; the same is true of run time except for the MM approach, which performs considerably better than the other three methods in this setting. In non-regularized composite quantile regression, however, differences between the methods in terms of estimation error are more apparent, as the IP method has larger average estimation error than the ADMM, MM, and CD approaches, while MM and CD are faster and ADMM slower than the IP algorithm. Comparisons between IP and ADMM methods for non-regularized quantile regression already exist in the literature \cite[Chapter 5]{HQR}. The results so far suggest that the MM approach is the best suited for non-regularized (composite) quantile regression among the four methods tested, especially for data sets with $p$ small relative to $n$. In regularized quantile regression, all of our approaches perform similarly in terms of variable selection, but CD and ADMM show clear superiority in run time, particularly relative to the IP and MM methods when $p$ is large. In the case of regularized composite quantile regression, CD and ADMM have run time superior to MM. Furthermore, MM shows a tendency to select irrelevant variables, likely due to the algorithm's matrix inversion and selection of an approximating parameter. This second set of results suggests that our CD approach is best suited for regularized (composite) quantile regression among the three methods tested, although care should be taken with regards to its theoretical convergence properties, as noted by \cite{earlyCD2}. In particular, since the penalty is not continuously differentiable in $\pmb\beta$ (so that the penalty is not separable as per \cite{earlyCD2}), convergence results do not apply. This situation is similar to that noted by \cite{cdproblem} in the context of fused lasso. In an example, CD is unable to achieve the global minimum of a strictly convex objective function. The authors show this problem stems from CD not allowing two particular components to be updated together, while no improvement to the value of the objective function is possible in one-component subproblem updates. With some specific modifications, however, \cite{cdproblem} show that this CD approach can be modified for highly competitive performance for the fused lasso problem.

Overall, our methods provide reliable and efficient algorithms to estimate solutions to quantile and composite quantile regression problems, including those regularized by an adaptive lasso penalty. Our methods, already implemented in the \texttt{cqrReg} package for R, widen the variety of algorithms available for quantile and composite quantile regression and greatly improve upon the run time of the existing advanced IP methods, particularly for large or high-dimensional data sets. Our ADMM method was competitive and is further amenable to parallelization, naturally lending itself to distributed computing to handle data that is both high-dimensional and extremely large in volume. ADMM may have future application in training deep neural networks through gains in estimation error and computation time. This is explored in greater depth by \cite{yulin} and \cite{admmhd} for big data and in sparse, high-dimensional settings.

\begin{acknowledgements}
Jueyu Gao acknowledges the supervision of Drs. Linglong Kong and Edit Gombay during his graduate studies. The authors have no declarations of interest to declare.
\end{acknowledgements}

\bibliographystyle{spbasic}      
\bibliography{cqrReg_references}   

\end{document}


\title{Supplementary Materials for\\Advanced Algorithms for Penalized Quantile and Composite Quantile Regression\thanks{Drs. Linglong Kong, Bei Jiang, and Di Niu are supported in part by the Natural Sciences and Engineering Research Council of Canada (NSERC).}
}


\author{Matthew Pietrosanu         \and
        Jueyu Gao \and
        Linglong Kong \and
        Bei Jiang \and
        Di Niu 
}


\institute{M. Pietrosanu, J. Gao, L. Kong, and B. Jiang \at
              Department of Mathematical \& Statistical Sciences, University of Alberta, Edmonton, AB, Canada \\
              Tel.: +780-492-3396\\
              Fax: +780-492-6826\\
              \email{pietrosa@ualberta.ca, jueyu@ualberta.ca, lkong@ualberta.ca, bei1@ualberta.ca}
              \and
          D. Niu \at
              Department of Electrical and Computer Engineering, University of Alberta, Edmonton, AB, Canada \\
              Tel.: + 780-492-3332 \\
              \email{dniu@ualberta.ca}
}


\maketitle

\appendix

\section{Quantile and Composite Quantile Regression Without Adaptive Lasso Regularization}
\label{app:theorem}

This supplementary appendix is structured as follows. Section \ref{appsubsec:qr} presents details of our approach to solving the standard quantile regression problem without regularization via alternating direction method of multipliers (ADMM), majorize-minimization (MM), and coordinate descent (CD) algorithms. For the sake of comparison, we also introduce a basic interior point (IP) approach. Section \ref{appsubsec:cqrNoReg} gives details on the generalization from quantile to composite quantile regression, again without regularization.

\subsection{Non-Regularized Quantile Regression} \label{appsubsec:qr}

The following Subsections \ref{appsubsec:qr:admm} through \ref{appsubsec:qr:cd} detail our approach to non-regularized quantile regression using the ADMM, MM, and CD algorithms. We place particular emphasis on the ADMM approach and first review its general setup. Subsection \ref{subsec:qrIP} introduces a basic IP method and a reformulation of the quantile regression problem accessible to the \texttt{Rmosek} optimization package for R \citep{Rmosek}. We use the notation presented in the main text throughout.

\subsubsection{Alternating Direction Method of Multipliers Algorithm} \label{appsubsec:qr:admm}

Before proceeding with an application to quantile regression, we review the general ADMM algorithm, which decomposes a given additively separable convex optimization problem into a number of sub-convex optimization problems. The general formulation of the ADMM problem is
\begin{align*}
\underset{\pmb x,\pmb z}{\min} \qquad & f(\pmb x)+g(\pmb z)\\
\text{subject to} \qquad & \pmb A\pmb x+\pmb B\pmb z = \pmb c,
\end{align*}
where $f$ and $g$ are convex, real-valued functions of $\pmb x$ and $\pmb z$, $\pmb A$ and $\pmb B$ are matrices, and $\pmb c$ is a constant vector. The augmented Lagrangian \citep{powell} of the above problem is written as $$L_\rho(\pmb x,\pmb z,\pmb y) = f(\pmb x) + g(\pmb z) + \pmb y^T(\pmb A\pmb x+\pmb B\pmb z-\pmb c) + \frac{\rho}{2}\lvert\lvert \pmb A\pmb x+\pmb B\pmb z-\pmb c\rvert\rvert^2_2,$$ where $\rho$ is a tuning parameter. Setting $u=\frac{1}{\rho}y$ and $u_k=\frac{1}{\rho}y_k$, we can obtain the (more convenient) scaled augmented Lagrangian $$L^\text{s}_\rho(\pmb x,\pmb z,\pmb y)=f(\pmb x)+g(\pmb z)+\frac{\rho}{2}\lvert\lvert \pmb A\pmb x+\pmb B\pmb z-\pmb c+\pmb u\rvert\rvert_2^2-\frac{\rho}{2}\pmb u_2.$$ The ADMM method optimizes the scaled augmented Lagrangian using the iterative scheme
\begin{align*}
\pmb x^{(t+1)} &= \underset{\pmb x}{\arg\min}\Big[f(\pmb x)+\frac{\rho}{2}\lvert\lvert \pmb A\pmb x+\pmb B\pmb z^{(t)}-\pmb c+\pmb u^{(t)}\rvert\rvert^2_2\Big],\\
\pmb z^{(t+1)} &= \underset{\pmb z}{\arg\min}\Big[g(\pmb z)+\frac{\rho}{2}\lvert\lvert \pmb A\pmb x^{(t+1)}+\pmb B\pmb z-\pmb c+\pmb u^{(t)}\rvert\rvert_2^2\Big],\\
\pmb u^{(t+1)} &= \pmb u^{(t)} + \pmb A\pmb x^{(t+1)}+\pmb B\pmb z^{(t+1)}-\pmb c.
\end{align*}
A generic stopping condition for the algorithm can be defined in terms of the primal and dual residuals, given by $\pmb r_\text{primal}^{(t+1)}=\pmb A\pmb x^{(t+1)}+\pmb B\pmb z^{(t+1)}-c$ and $\pmb r_\text{dual}^{(t+1)}=\rho \pmb A^T\pmb B(\pmb z^{(t+1)}-\pmb z^{(t)})$. The program can be made to terminate if both
\begin{align*}
\lvert\lvert \pmb r_\text{primal}^{(t)}\rvert\rvert_2 &\leq \varepsilon_\text{primal} = \sqrt{p}\varepsilon_\text{abs} + \varepsilon_\text{rel}\max\{\lvert\lvert \pmb A\pmb x^{(t)}\rvert\rvert_2,\lvert\lvert \pmb B\pmb z^{(t)}\rvert\rvert_2, \lvert\lvert \pmb c\rvert\rvert_2\}\\
\lvert\lvert \pmb r_\text{dual}^{(t)}\rvert\rvert_2 &\leq \varepsilon_\text{dual} = \sqrt{n}\varepsilon_\text{abs} + \varepsilon_\text{rel}\lvert\lvert \pmb A^T\pmb y^{(t)}\rvert\rvert_2,
\end{align*}
where $p$ and $n$ are the length of $c$ and $\pmb A^T\pmb y^{(t)}$, respectively. In our applications, we set $\varepsilon_\text{abs}=10^{-2}$ and $\varepsilon_\text{rel}=10^{-4}$.

We apply the ADMM algorithm \citep{ADMMnew1} by reformulating quantile regression as the convex optimization problem
\begin{align*}
\underset{\pmb \beta\in\mathbb{R}^{p+1}}{\min}\qquad&\sum_{i=1}^n\rho_\tau(r_i)\\
\text{subject to }\qquad&  \pmb X\pmb \beta+\pmb r=\pmb Y,
\end{align*}
where $\pmb r$ is a vector of residuals. The intercept term is accounted for in both $\pmb \beta$ and $\pmb X$. Using the general procedure of ADMM, taking $f=0$ and $g$ as a function of $\pmb r$ to be the entire objective function, we obtain the iterative scheme
\begin{align*}
\pmb \beta^{(t+1)} &= \underset{\pmb \beta\in\mathbb{R}^{p+1}}{\arg\min}~~\frac{\rho}{2}||\pmb Y-\pmb r^{(t)}-\pmb X\pmb \beta+\pmb u^{(t)}/\rho||_2^2,\\
\pmb r^{(t+1)} &= \underset{\pmb r\in\mathbb{R}^n}{\arg\min}~~\sum_{i=1}^n\rho_\tau(r_i)+\frac{\rho}{2}||\pmb Y-\pmb r-\pmb X\pmb \beta^{(t+1)}+\pmb u^{(t)}/\rho||_2^2,\\
\pmb u^{(t+1)} &= \pmb u^{(t)} + \rho(\pmb Y-\pmb r^{(t+1)}-\pmb X\pmb \beta^{(t+1)}),
\end{align*}
where $\pmb u$ is the rescaled Lagrange multiplier and $\rho>0$ is a penalty parameter. The update for $\pmb r$ can be written in a closed form as $S_{1/\rho}\big(\pmb c-(2\pmb \tau_{n\times1}-\pmb 1_{n\times1})/\rho\big)$, where $\pmb c=\pmb Y-\pmb X\pmb \beta^{(t)}+\pmb u^{(t)}/\rho$ and, for real $a$, the function $S_a:\mathbb{R}^m\rightarrow\mathbb{R}^m$ is defined component-wise via $(S_a(\pmb v))_i = (v_i-a)_+-(-v_i-a)_+$. The closed form for the update of $\pmb \beta$ is given by $(\pmb X^T\pmb X)^{-1}\pmb X^T(\pmb Y-\pmb r^{(t)}+\pmb u^{(t)}/\rho)$. For reference, $\rho$ is chosen to be 1.2 by \cite{ADMMnew1}. In the quantile regression setting, we have that
\begin{align*}
\pmb r_\text{primal}^{(t+1)} &= \pmb Y-\pmb X\pmb \beta^{(t+1)}-\pmb r^{(t+1)},\\
\pmb r_\text{dual}^{(t+1)} &= \rho \pmb X^T(\pmb r^{(t+1)}-\pmb r^{(t)}),\\
\varepsilon_\text{primal} &= \sqrt{n}\varepsilon_\text{abs} + \varepsilon_\text{rel}\max\{\lvert\lvert \pmb X\beta^{(t+1)}\rvert\rvert_2^2,\lvert\lvert \pmb r^{(t+1)}\rvert\rvert_2^2,\lvert\lvert \pmb Y\rvert\rvert_2^2\},\\
\varepsilon_\text{dual} &= \sqrt{p}\varepsilon_\text{abs}+ \varepsilon_\text{rel}\lvert\lvert \pmb X^T\pmb u^{(t+1)}\rvert\rvert_2^2.
\end{align*}

\subsubsection{Majorize-Minimization Algorithm} \label{appsubsec:qr:mm}

We use the MM algorithm developed by \cite{MM} and \cite{MMR} to solve the quantile regression problem without regularization. Our approach is exactly the same as in the main text, but we instead ignore the majorization of the penalty term in the quantile regression objective function. Construct a function $\rho_\tau^\varepsilon(r)$ based on some perturbation parameter $\varepsilon>0$ that will be used to approximate the quantile regression objective function $L(\pmb \beta)$. For any residual $r$, define $\rho_\tau^\varepsilon(r) = \rho_\tau(r) - \frac{\varepsilon}{2}\ln(\varepsilon+|r|)$, and the subsequent approximation of $L(\pmb \beta)$ by $L^\varepsilon(\pmb \beta) = \sum_{i=1}^n\rho_\tau^\varepsilon(r_i)$. At the $t$-th iteration of the algorithm, for each current residual value $r_i^{(t)}=r_i^{(t)}(\pmb \beta^{(t)})$, $\rho_\tau^\varepsilon(r)$ is majorized by the quadratic function $$\xi_\tau^\varepsilon(r|r_i^{(t)}) = \frac{1}{4}\Biggl[\frac{r^2}{\varepsilon+|r_i^{(t)}|}+(4\tau-2)r+c\Biggr],$$ for some solvable constant $c$ that satisfies the equation $\xi(r_i^{(t)}|r^{(t)})=\rho_\tau^\varepsilon(r^{(t)})$. The MM algorithm minimizes the majorizer of $L^\varepsilon(\pmb \beta)$, namely, $$Q^\varepsilon(\pmb \beta|\pmb \beta^{(t)})=\sum_{i=1}^n\xi_\tau^\varepsilon(r_i|r_i^{(t)}),$$ with the argument minimum taken as the updated value $\pmb \beta^{(t+1)}$ of $\pmb \beta$. For the $t$-th iteration of the algorithm, given an updated value $\pmb \beta^{(t)}$ for $\pmb \beta$, we generate and minimize a new majorized quadratic function $Q^\varepsilon(\cdot|\pmb \beta^{(t)})$ and implement a Newton-Raphson iterative method to obtain an updated value $\pmb \beta^{(t+1)}$ for $\pmb \beta$.

\subsubsection{Coordinate Descent Algorithm} \label{appsubsec:qr:cd}

To implement quantile regression, we use an extended version of the greedy CD method put forward by Edgeworth and, more recently, further developed by \cite{CD}. In each iteration, for fixed $\pmb \beta\in\mathbb{R}^p$, replace $b_0$ by the $\tau$-th sample quantile of the residuals $y_i-\pmb X_i^T\pmb \beta$ for $i=1,\dots,n$: this will necessarily decrease the value of the objective function. Define $\Theta_i = \rho_\tau(r_i)$ for $i=1,\dots,n$. For each element $\beta_m$ for $m=1,\dots,p$ of $\pmb \beta$, rewrite the loss function as $$L(b_0,\pmb \beta)=L_{m}(b_0,\pmb \beta) = \sum_{i=1}^n\lvert x_{im}\rvert\Biggl\lvert\frac{y_i-b_0-\sum_{j=1,\,j\neq m}^px_{ij}\beta_j}{x_{im}}-\beta_m\Biggr\rvert\cdot\Theta_i,$$ so that the CD algorithm applies. For each fixed $m$, sort the values of $$z_i=\frac{y_i-b_0-\sum_{j=1,\,j\neq m}^px_{ij}\beta_j}{x_{im}}$$ for $i=1,\dots,n$ and update $\beta_m$ to be the $i^*$-th order statistic $z_{(i^*)}$ satisfying both
$$\sum_{j=1}^{i^*-1}w_{(j)} < \frac{1}{2}\sum_{j=1}^n w_{(j)}
\qquad\text{and}\qquad
\sum_{j=1}^{i^*}w_{(j)} \geq \frac{1}{2}\sum_{j=1}^n w_{(j)},$$
where $w_i=\lvert x_{im}\rvert\cdot\Theta_i$. In other words, using the weights $w_i$, the selected $z_{(i^*)}$ is the weighted median of all $z_i$ (for the fixed value of $m$). At the end of each iteration, we check for the convergence of $\pmb \beta$ and stop the algorithm using an absolute value difference threshold of $10^{-3}$.

\subsubsection{Interior Point Algorithm} \label{subsec:qrIP}

Interior point (IP) methods generally reach an optimal solution by travelling within rather than on the boundary of the feasible set. Though studied as early as the 1950s and 1960s, IP methods arguably first gained widespread interest with the landmark paper by \cite{landmarkIP}, who proposed an efficient, polynomial time IP algorithm for linear programs with performance rivalling the existing simplex method. \cite{nesterovIP} later extended these results to a range of convex optimization problems while maintaining polynomial time. In the present day, advanced IP methods and code for both linear and non-linear programs are widely available and well-studied in the literature \citep{IPtextbook}. IP algorithms have also received considerable attention and success in applications to non-linear, non-convex optimization problems \citep{nonIP}.

We can implement quantile regression using an IP algorithm by reformulating the optimization problem as a linear program and making use of existing optimization packages such as \texttt{Rmosek} \citep{Rmosek}.
\texttt{Rmosek} can implement an IP algorithm to solve problems of the form
\begin{align*}
\underset{\pmb x\in\mathbb{R}^n}{\min}\qquad &\pmb c^T\pmb x + c_0\\
\text{subject to}\qquad &\pmb l^ c\leq \pmb A\pmb x\leq \pmb u^c\\
&\pmb l^x\leq x \leq \pmb u^x,
\end{align*}
where $\pmb A\in\mathbb{R}^{m\times n}$ is a constraint matrix; $\pmb c\in\mathbb{R}^n$ and $c_0\in\mathbb{R}$ the objective function coefficients and constant; $\pmb l^c,\pmb u^c\in\mathbb{R}^m$ the lower and upper constraint bounds; and $\pmb l^x,\pmb u^x\in\mathbb{R}^n$ the lower and upper variable bounds. For notational simplicity, $\leq$ is taken to mean component-wise comparison of vectors. Alternatively, other R packages such as \texttt{quantreg} exist specifically for quantile regression and make use of IP methods. The IP approach for quantile regression in \texttt{quantreg} is based on the method of \cite{hare}, with recent modifications including the prediction-correction algorithm of \cite{mehrotra}. Lasso penalized quantile regression in \texttt{quantreg} uses a Frisch-Newton method.

Let $\pmb u,\pmb v\in\mathbb{R}^n_{\geq0}$ be a vector of the positive and negative parts, respectively, of the residuals $\pmb r=(r_1,\dots,r_n)$, and $\pmb \beta\in\mathbb{R}^{p+1}$ a vector of parameters including the intercept. The quantile regression problem without regularization can be formulated for use in existing IP optimization routines such as \texttt{Rmosek} via
\begin{align*}
\underset{\pmb \beta\in\mathbb{R}^{p+1},~\pmb u,\pmb v\in\mathbb{R}^n}{\min} \qquad &\tau \pmb 1_{n\times1}^T\pmb u + (1-\tau)\pmb 1_{n\times1}^T\pmb v\\
\text{subject to} \qquad & \pmb Y=\pmb X\pmb \beta+\pmb u-\pmb v\\
&\pmb 0_{n\times1} \leq \pmb u \leq \pmb \infty_{n\times1}\\
&\pmb 0_{n\times1} \leq \pmb v \leq \pmb \infty_{n\times1}.
\end{align*}

As an aside, to incorporate an adaptive lasso penalty into the problem, we can rewrite the problem as a linear program accessible to existing IP routines via\begin{align*}
\underset{\pmb \beta\in\mathbb{R}^{p+1},~\pmb u,\pmb v\in\mathbb{R}^n}{\min} \qquad &\tau \pmb 1_{n\times1}^T\pmb u + (1-\tau)\pmb 1_{n\times1}^T\pmb v+p_\lambda(\lvert\pmb \beta\rvert)\\
\text{subject to} \qquad & \pmb \beta \leq \pmb \beta^*\\
& -\pmb \beta \leq \pmb \beta^*\\
& \pmb 0_{n\times1} \leq \pmb \beta^* \leq \pmb \infty_{n\times1}\\
& \pmb 0_{n\times1} \leq \pmb u \leq \pmb \infty_{n\times1}\\
& \pmb 0_{n\times1} \leq \pmb v \leq \pmb \infty_{n\times1}.
\end{align*}

\subsection{Composite Quantile Regression} \label{appsubsec:cqrNoReg}

This section shows details of the extension from quantile to composite quantile regression without regularization. Subsections \ref{appsubsec:cqr:admm}, \ref{appsubsec:cqr:mm}, and \ref{appsubsec:cqr:cd} extend the above non-regularized quantile regression procedures using ADMM, MM, and CD algorithms, respectively. Subsection \ref{appsubsec:cqr:ip} formulates the problem for use in \texttt{Rmosek} \citep{Rmosek} or other IP methods for linear programs. We use the notation presented in the main text throughout.

\subsubsection{Alternating Direction Method of Multipliers Algorithm} \label{appsubsec:cqr:admm}

Written in the ADMM form, the composite quantile regression problem can be expressed as
\begin{align*}
\underset{\beta\in\mathbb{R}^{p+1}}{\min}\qquad&\sum_{k=1}^K\sum_{i=1}^n\rho_{\tau_k}(r_{ik})\\
\text{subject to }\qquad&  \pmb X^*\pmb \beta+\pmb r=\pmb Y^*,
\end{align*}
where we assume that the intercept term is accounted for in both $\pmb \beta$ and $\pmb X$. The ADMM approach is applied in exactly the same way as in Subsection \ref{appsubsec:qr:admm}, yielding the iterative update scheme
\begin{align*}
\pmb \beta^{(t+1)} &= \underset{\pmb \beta\in\mathbb{R}^{p+K}}{\arg\min}~~\frac{\rho}{2}||\pmb Y^*-\pmb r^{(t)}-\pmb X^*\pmb \beta+\pmb u^{(t)}/\rho||_2^2,\\
\pmb r^{(t+1)} &= \underset{\pmb r\in\mathbb{R}^{nK}}{\arg\min}~~\sum_{k=1}^K\sum_{i=1}^n\rho_{\tau_k}(r_{ik})+\frac{\rho}{2}||\pmb Y^*-\pmb r-\pmb X^*\pmb \beta^{(t+1)}+\pmb u^{(t)}/\rho||_2^2,\\
\pmb u^{(t+1)} &= \pmb u^{(t)} + \rho(\pmb Y^*-\pmb r^{(t+1)}-\pmb X^*\pmb \beta^{(t+1)}),
\end{align*}
where $\pmb c=\pmb Y^*-\pmb X^*\pmb \beta^{(t)}+\pmb u^{(t)}/\rho$; and residuals
\begin{align*}
\pmb r_\text{primal}^{(t+1)} &= \pmb Y^*-\pmb X^*\pmb \beta^{(t+1)}-\pmb r^{(t+1)},\\
\pmb r_\text{dual}^{(t+1)} &= \rho {\pmb X^*}^T(\pmb r^{(t+1)}-\pmb r^{(t)}),\\
\varepsilon_\text{primal} &= \sqrt{n}\varepsilon_\text{abs} + \varepsilon_\text{rel}\max\{\lvert\lvert \pmb X^*\pmb \beta^{(t+1)}\rvert\rvert_2^2,\lvert\lvert \pmb r^{(t+1)}\rvert\rvert_2^2,\lvert\lvert \pmb Y^*\rvert\rvert_2^2\},\\
\varepsilon_\text{dual} &= \sqrt{p}\varepsilon_\text{abs}+ \varepsilon_\text{rel}\lvert\lvert {\pmb X^*}^T\pmb u^{(t+1)}\rvert\rvert_2^2.
\end{align*}
A generic stopping condition requiring $\lvert\lvert \pmb r_\text{primal}^{(t)}\rvert\rvert\leq\varepsilon_\text{primal}$ and $\lvert\lvert \pmb r_\text{dual}^{(t)}\rvert\rvert\leq\varepsilon_\text{dual}$ for termination can be imposed. We again take $\pmb u$ as the rescaled Lagrange multiplier and $\rho>0$ as a penalty parameter. Generalizing from quantile regression, the update for $\pmb r$ can be written in a closed form as $S_{1/\rho}\big(\pmb c-(2\pmb \tau^*-\pmb 1_{n\times1})/{\rho}\big)$, with $S_a$ as defined previously for real $a$. The closed form update for $\pmb \beta$ is given by $({\pmb X^*}^T\pmb X^*)^{-1}{\pmb X^*}^T(\pmb Y^*-\pmb r^{(t)}+\pmb u^{(t)}/\rho)$.

\subsubsection{Majorize-Minimization Algorithm} \label{appsubsec:cqr:mm}

An extension of the MM algorithm from quantile to composite quantile regression simply involves the incorporation of additional quantile levels. We use the same function $\rho_\tau^\varepsilon(r) = \rho_\tau(r) - \frac{\varepsilon}{2}\ln(\varepsilon+|r|)$ to approximate the composite quantile regression objective function via $L^\varepsilon(\pmb \beta) = \sum_{k=1}^K\sum_{i=1}^n\rho_{\tau_k}^\varepsilon(r_{ik})$. We also use the same function $\xi$ as defined in Subsection \ref{appsubsec:qr:mm} to majorize $\rho_\tau^\varepsilon$. At the $t$-th iteration of the algorithm, for each current residual value $r_{ik}^{(t)}=r_{ik}^{(t)}(\pmb \beta^{(t)})$, we have that $\rho_{\tau_k}^\varepsilon(r)$ is majorized by the quadratic function $$\xi_{\tau_k}^\varepsilon(r|r_{ik}^{(t)}) = \frac{1}{4}\Biggl[\frac{r^2}{\varepsilon+|r_{ik}^{(t)}|}+(4\tau_k-2)r+c\Biggr],$$ for some solvable constant $c$ that satisfies the equation $\xi(r_{ik}^{(t)}|r_{ik}^{(t)})=\rho_{\tau_k}^\varepsilon(r_{ik}^{(t)})$. The MM algorithm minimizes the majorizer of $L^\varepsilon(\pmb \beta)$, namely, $$Q^\varepsilon(\pmb \beta|\pmb \beta^{(t)})=\sum_{k=1}^K\sum_{i=1}^n\xi_{\tau_k}^\varepsilon(r_{ik}|r_{ik}^{(t)}),$$ with the argument minimum taken as the updated value $\pmb \beta^{(t+1)}$ of $\pmb \beta$. In practice, for the $t$-th iteration of the algorithm, given an updated value $\pmb \beta^{(t)}$ for $\pmb \beta$, we generate and minimize a new majorized quadratic function $Q^\varepsilon(\cdot|\pmb \beta^{(t)})$ using a Newton-Raphson iterative method. The argument minimum is taken as the updated value $\pmb \beta^{(t+1)}$ for $\pmb \beta$.

\subsubsection{Coordinate Descent Algorithm} \label{appsubsec:cqr:cd}

To apply the CD method to composite quantile regression, we rewrite the composite quantile regression objective function in the required CD form. For any $m=1,\dots,p$, we have $$L_m(b_1,\dots,b_k,\pmb \beta) = \sum_{k=1}^K\sum_{i=1}^n\lvert x_{im}\rvert\Biggl\lvert\frac{y_i-b_k-\sum_{j=1,\,j\neq m}^px_{ij}\beta_j}{x_{im}}-\beta_m\Biggr\rvert\cdot\Theta_{ik},$$ with $\Theta_{ik} = \rho_{\tau_k}(r_{ik})$ for $i=1,\dots,n$ and $k=1,\dots,K$. In each iteration, and for fixed $\pmb \beta$, replace $b_k$, for $k=1,\dots,K$, with the $\tau$-th sample quantile of the residuals $y_i -\pmb X_i^T\pmb \beta$ for $i=1,\dots,n$. To update $\beta_m$ for $m=1,\dots,p$, sort the numbers $$z_{ik} = \frac{y_i-b_k-\sum_{j=1,\,j\neq m}^px_{ij}\beta_j}{x_{im}},$$ for $i=1,\dots,n$ and $k=1,\dots,K$. Update $\beta_m$ with the value of the $i^*$-th order statistic $z_{(i^*)}$ satisfying both
$$\sum_{j=1}^{i^*-1}w_{(j)} < \frac{1}{2}\sum_{j=1}^{nK} w_{(j)}
\qquad\text{and}\qquad
\sum_{j=1}^{i^*}w_{(j)} \geq \frac{1}{2}\sum_{j=1}^{nK} w_{(j)},$$
where $w_{ik}=\lvert x_{im}\rvert\cdot\Theta_{ik}$. At the end of each iteration, we check for the convergence of $\pmb \beta$ and stop the algorithm using an absolute value difference threshold of $10^{-3}$.

\subsubsection{Interior Point Algorithm} \label{appsubsec:cqr:ip}

The extension of the previous IP method from quantile to composite quantile regression simply requires us to account for the extra quantile levels in the objective function and the resulting extra residuals. The problem can be formulated as a linear program via
\begin{align*}
\underset{\pmb \beta\in\mathbb{R}^{p+K},~\pmb u_k,\pmb v_k\in\mathbb{R}^n}{\min} \qquad & \sum_{k=1}^K \tau_k\pmb 1_{n\times1}^T\pmb u_k + (1-\tau_k)\pmb 1_{n\times1}^T\pmb v_k\\
\text{subject to} \qquad & \pmb Y=\pmb X\pmb \beta+\pmb u_k-\pmb v_k\\
& \pmb 0 \leq \pmb u_k \leq \pmb \infty\\
& \pmb 0 \leq \pmb v_k \leq \pmb \infty,
\end{align*}
where each constraint is to hold for all $k=1,\dots,K$.

\renewcommand{\thesubsection}{B.\arabic{subsection}}
\section{Composite Quantile Regression with Adaptive Lasso Regularization} \label{appsubsec:cqrreg}

Here we give explicit details regarding the ADMM, MM, and CD methods for composite quantile regression with adaptive lasso regularization. An IP approach is also given for comparison.

\subsection{Alternating Direction Method of Multipliers Algorithm}

Applying ADMM in the composite quantile setting with adaptive lasso regularization, we obtain the iterative update scheme
\begin{align*}
\pmb r^{(t+1)} &= \underset{\pmb r\in\mathbb{R}^{nK}}{\arg\min}~~\sum_{k=1}^K\sum_{i=1}^n\rho_{\tau_k}(r_{ik})+\frac{\rho}{2}||\pmb Y^*-\pmb r-\pmb X^*\pmb \beta^{(t)}+\pmb u^{(t)}/\rho||_2^2,\\
\pmb \beta^{(t+1)} &= \underset{\pmb \beta\in\mathbb{R}^{p+K}}{\arg\min}~~\frac{\rho}{2}||\pmb Y^*-\pmb r^{(t+1)}-\pmb X^*\pmb \beta+\pmb u^{(t)}/\rho||_2^2+\lambda\sum_{j=1}^p\sfrac{\lvert{\beta}_j\rvert}{\lvert\beta_j^\text{CQR}\rvert^2},\\
\pmb u^{(t+1)} &= \pmb u^{(t)} + \rho(\pmb Y^*-\pmb r^{(t+1)}-\pmb X^*\pmb \beta^{(t+1)}),
\end{align*}
where $\pmb c=\pmb Y^*-\pmb X^*\pmb \beta^{(t)}+\pmb u^{(t)}/\rho$; and residuals
\begin{align*}
\pmb r_\text{primal}^{(t+1)} &= \pmb Y^* -\pmb X^*\pmb \beta^{(t+1)}- \pmb r^{(t+1)},\\
\pmb r_\text{dual}^{(t+1)} &= \rho {\pmb X_*^*}^T(\pmb r^{(t+1)}-\pmb r^{(t)}),\\
\varepsilon_\text{primal} &= \sqrt{n}\varepsilon_\text{abs} + \varepsilon_\text{rel}\max\{\lvert\lvert \pmb X_*^*\pmb \beta_*^{(t+1)}\rvert\rvert_2^2, \lvert\lvert \pmb r^{(t+1)}\rvert\rvert_2^2, \lvert\lvert \pmb b^*-\pmb Y \rvert\rvert_2^2\},\\
\varepsilon_\text{dual} &= \sqrt{p}\varepsilon_\text{abs}+\varepsilon_\text{rel}\lvert\lvert {\pmb X^*}^T\pmb u^{(t+1)}\rvert\rvert_2^2.
\end{align*}
We again take $\pmb u$ as the rescaled Lagrange multiplier and $\rho>0$ as a penalty parameter. As before, the update for $\pmb r$ can be written in a closed form as $S_{1/\rho}\big(\pmb c-(2\pmb \tau^*-\pmb 1_{n\times1})/{\rho}\big)$, with $S_a$ as defined previously for real $a$. With adaptive lasso regularization, the update for $\pmb \beta$ does not have a closed form but can be viewed as a least squares optimization problem with adaptive lasso penalty. We implement existing numerical methods to solve this problem and update $\pmb \beta$.

\subsection{Majorize-Minimization Algorithm}

An extension of the MM method for adaptive lasso regularized quantile regression to regularized composite quantile regression involves a minor change to incorporate multiple quantile levels into the majorized objective function. Using the same function $\rho_\tau^\varepsilon(r) = \rho_\tau(r)-\frac{\varepsilon}{2}\ln(\varepsilon+|r|)$ as before with perturbation parameter $\varepsilon>0$ to approximate $\rho_\tau(r)$, we can approximate the regularized quantile regression objective function via $$\sum_{k=1}^K\sum_{i=1}^n\rho_{\tau_k}^\varepsilon(r_{ik})++\lambda\sum_{j=1}^p\frac{1}{\lvert\beta^\text{CQR}_j\rvert^2}\Biggl[|\beta^{(t)}_j|+ \frac{\big(\beta_j^2-(\beta^{(t)}_j)^2\big)\mathrm{sgn}(\beta_j^{(t)})}{2|\beta^{(t)}_j+\varepsilon|}\Biggr].$$ Define, as before, $$\xi_{\tau_k}^\varepsilon(r|r_{ik}^{(t)}) = \frac{1}{4}\Biggl[\frac{r^2}{\varepsilon+|r_{ik}^{(t)}|}+(4\tau_k-2)r+c\Biggr].$$ In the $t$-th iteration of the MM algorithm, the (approximated) objective function is majorized by $$Q^\varepsilon(\pmb \beta|\pmb \beta^{(t)}) = \sum_{k=1}^K\sum_{i=1}^n\xi_{\tau_k}^\varepsilon(r_{ik}|r_{ik}^{(t)})+\lambda\sum_{j=1}^p\frac{1}{\lvert\beta_j^\text{CQR}\rvert^2}\Biggl[|\beta^{(t)}_j|+ \frac{\big(\beta_j^2-(\beta^{(t)}_j)^2\big)\mathrm{sgn}(\beta_j^{(t)})}{2|\beta^{(t)}_j+\varepsilon|}\Biggr].$$ Given an updated value $\pmb \beta^{(t)}$ for $\pmb \beta$, we generate a new majorizing function $Q^\varepsilon(\cdot|\pmb \beta^{(t)})$ and implement a Gauss-Newton iterative method to estimate and update the value of $\pmb \beta$.

\subsection{Coordinate Descent Algorithm}

As discussed in the main text, the CD method for regularized composite quantile regression simply adjusts the objective function to account for the extra quantile levels as $$L_m(b_1,\dots,b_k,\pmb \beta) = \sum_{k=1}^K\sum_{i=1}^n\lvert x_{im}\rvert\Biggl\lvert\frac{y_i-b_k-\sum_{j=1,\,j\neq m}^px_{ij}\beta_j}{x_{im}}-\beta_m\Biggr\rvert\cdot\Theta_{ik}+p_\lambda(\lvert\pmb \beta\rvert).$$ In each iteration, for $k=1,\dots,K$, replace each $b_k$ with the $\tau$-th sample quantile of the residuals $y_i-\pmb X_i^T\pmb \beta$ for $i=1,\dots,n$. Define $z_{ik}= \frac{1}{x_{im}}\big(y_i-b_k-\sum_{j=1,\,j\neq m}^px_{ij}\beta_j\big)$ if $r_{ik}\geq0$ and $z_{ik}=0$ if $r_{ik}<0$. Update $\beta_m$ to the value of the $i^*$-th order statistic $z_{(i^*)}$ satisfying both
$$\sum_{j=1}^{i^*-1}w_{(j)} < \frac{1}{2}\sum_{j=1}^{nK} w_{(j)}
\qquad\text{and}\qquad
\sum_{j=1}^{i^*}w_{(j)} \geq \frac{1}{2}\sum_{j=1}^{nK} w_{(j)},$$
where $w_{ik}=\lvert x_{im}\rvert\cdot\Theta_{ik}$ if $r_{ik}\geq0$ and $w_{ik}=\sfrac{\lambda}{\lvert\beta_m^\text{CQR}\rvert^2}$ if $r_{ik}<0$. At the end of each iteration, check for the convergence of $\pmb \beta$ and stop the algorithm using an absolute value difference threshold of $10^{-3}$.

\subsection{Interior Point Algorithm}

Adaptive lasso regularized composite quantile regression is formulated by incorporating an appropriate penalty term into the linear program of Subsection \ref{appsubsec:cqr:ip}. This form is appropriate for the IP implementation in the \texttt{Rmosek} package \citep{Rmosek} and is given by
\begin{align*}
\underset{\pmb \beta\in\mathbb{R}^{p+K},~\pmb u_k,\pmb v_k\in\mathbb{R}^{n}}{\min} \qquad & \sum_{k=1}^K \tau_k\pmb 1_{n\times1}^T\pmb u_k + (1-\tau_k)\pmb 1_{n\times1}^T\pmb v_k+p_\lambda(\lvert\pmb \beta\rvert)\\
\text{subject to} \qquad & \pmb Y=\pmb X\pmb \beta+\pmb u_k-\pmb v_k\\
& \pmb \beta\leq\pmb \beta^*\\
& -\pmb \beta\leq\pmb \beta^*\\
& \pmb 0\leq \pmb u_k \leq\pmb \infty\\
& \pmb 0\leq \pmb v_k \leq\pmb \infty,\\
\end{align*}
where constraints are to hold for all $k=1,\dots,K$.

\begin{acknowledgements}
Jueyu Gao acknowledges the supervision of Drs. Linglong Kong and Edit Gombay during his graduate studies. The authors have no declarations of interest to declare.
\end{acknowledgements}

\bibliographystyle{spbasic}      
\bibliography{cqrReg_references}   